\documentclass[11pt]{article}
\usepackage{graphicx}

\def\singleandabitspaced{\baselineskip=\normalbaselineskip\multiply
    \baselineskip by 110\divide\baselineskip by 100}
\def\singlespaced{\baselineskip=\normalbaselineskip}

\newcommand{\centeron}[2]{{\setbox0=\hbox{#1}\setbox1=\hbox{#2}\ifdim
                             \wd1>\wd0\kern.5\wd1\kern-.5\wd0\fi \copy0
                             \kern-.5\wd0\kern-.5\wd1\copy1\ifdim\wd0>\wd1
                             \kern.5\wd0\kern-.5\wd1\fi}}
\newcommand{\ltap}{\>\centeron{\raise.35ex\hbox{$<$}}
                     {\lower.65ex\hbox{$\sim$}}\>}
\newcommand{\gtap}{\>\centeron{\raise.35ex\hbox{$>$}}
                     {\lower.65ex\hbox{$\sim$}}\>}

\begin{document}

\singlespaced

\begin{titlepage}

\vspace*{-16mm}
\begin{flushright}
{\small
MADPH-05-1415 \\
hep-ph/0502449  \\
}
\end{flushright}

\begin{center}
\vspace*{0.8in}
\mbox{\Large \textbf{High Energy Neutrinos from the TeV Blazar 1ES 1959+650}} \\
\vspace*{1.6cm}
{\large Francis Halzen$^1$ and Dan Hooper$^2$} \\
\vspace*{0.5cm}
{$^1$ University of Wisconsin, 1150 University Ave., Madison, WI 53706, USA \\
$^2$ University of Oxford, 
Denys Wilkinson Building, Keble Road, Oxford, OX1-3RH, UK} \\
\vspace*{0.6cm}
{\tt halzen@pheno.physics.wisc.edu, hooper@astro.ox.ac.uk} \\
\vspace*{1.5cm}
\end{center}

\begin{abstract} 

The AMANDA neutrino telescope has recently reported the detection of high-energy neutrinos spatially and temporally coincident with the flaring of the TeV blazar 1ES 1959+650. At present, the statistical significance of this observation cannot be reliably assessed, however. In this letter, we investigate whether circumstances exist where the source can produce the flux implied by the coincident events. We show that if the TeV gamma-ray emission observed from  1ES 1959+650 or other nearby TeV blazars is the result of accelerated protons interacting with nucleons, it is reasonable that AMANDA could detect several events during a flaring period. Such rates require that the spectral index of the source be rather high (for instance $\sim2.8$ for 1ES 1959+650) and that the Lorentz factor of the jet be fairly small ($\Gamma \sim 1$).

\end{abstract}

\end{titlepage}

\newpage

\singleandabitspaced

\section{Introduction}

Blazars, the class of Active Galactic Nuclei (AGN) with collimated jets aligned along the direction of observation, have been detected over a wide range of wavelengths ranging from radio to very high energy gamma-rays. Currently, six of these objects have been identified as TeV emitting blazars \cite{mrk421,mrk501,pks,1es2344,1h,1es1959}. These objects are, cosmologically speaking, fairly local (between 140 and 600 Mpc) and extremely luminous, particularly during periods of flaring activity ($\sim 10^{45}-10^{49}$ erg/s, inferred isotropically). Such flares may extend for hours or days at a time \cite{Mukherjee:2000nf,Sikora:2001ze,Catanese:1999cn}.

Blazars typically display spectra with enhanced emission over two energy ranges: the IR/X-ray and MeV/TeV peaks. The lower energy peak is generally agreed to be the product of synchrotron radiation from accelerated electrons \cite{Sikora:2001ze,Mukherjee:1999kw,Sreekumar:1999xw,Boettcher:1999xn}. The origin of the higher energy peak is not yet agreed upon, however. In leptonic models, inverse Compton scattering of synchrotron photons (or other ambient photons) by accelerated electrons generates this high energy emission. In hadronic models \cite{Mannheim:1993jg,Nellen:1992dw,Mucke:2000rn,Aharonian:2000pv,Halzen:2002pg}, on the other hand, MeV-TeV gamma-rays are produced through proton collisions with radiation or gas clouds surrounding the object. These collisions generate neutral and charged pions which decay, producing very high energy gamma-rays and (yet unobserved) neutrinos. The observation of high energy neutrinos would be a strong confirmation of the hadronic blazar model. Although roughly 60 blazars have been observed in the MeV-GeV range, only 6 have been detected at TeV energies.

Of the six known TeV blazars, the Markanians are the most nearby ($z=0.031$ and 0.033, respectively).  1ES 2344+514 and 1ES 1959+650 are not much further ($z=0.044$ and 0.047), while PKS 2155-304 and 1H 1426+428 are the most distant ($z=0.12$ and 0.13). All of these except for PKS 2155-304 are northern hemisphere objects, which is necessary to avoid the atmospheric neutrino background in an Antarctic neutrino telescope such as AMANDA or IceCube (PKS 2155-304 could be a potential source for Antares, however).

During the spring and summer of 2002, 1ES 1959+650 underwent a period of bright flaring activity. In May of that year, flaring was observed in the TeV range by the Whipple \cite{whipple} and HEGRA \cite{hegra} experiments as well as in the X-ray range by the RXTE experiment. The X-ray observations of 1ES 1959+650 then showed a gradual decline throughout the following month. During that same month, however, Whipple observed the recurrence of flaring in the TeV. The observation of this ``orphan'' TeV flare (with no X-ray counterpart) is in striking disagreement with the predictions of the leptonic blazar model. In leptonic models, flaring at TeV energies is generally accompanied by simultaneous flaring in the lower energy peak \footnote{For arguments attempting to reconcile 1ES 1959+650 with leptonic blazar models, see Ref.~\cite{krawczynski}}. This argument designates 1ES 1959+650 as a TeV blazar likely to generate gamma-rays hadronically \cite{Bottcher:2004qs}, and therefore likely to generate a substantial flux of high energy neutrinos.

Very recently, the AMANDA collaboration has reported the detection of two neutrinos coincident with TeV flares seen by Whipple from the direction of 1ES 1959+650 \cite{observation1,observation2}. These events were not uncovered in a blind analysis, however, and therefore their statistical significance cannot be reliably estimated. The AMANDA collaboration has adopted the position that a potential signal, studied after the data has been unblinded, can no longer be statistically evaluated.

Although the probability of background events occurring coincidentally with the orphan flare of 1ES 1959+650 is quite small ($\sim10^{-3}$), we emphasize that it is impossible to treat this statistically, as there is no known number of trials to dilute the result by. With this in mind, we proceed to explore the implications of these events, in the case that they are the product of hadronic interactions in the jet of 1ES 1959+650.

Our conclusion is that the events observed from the direction of 1ES 1959+650 by the AMANDA experiment are consistent with the model of proton-proton collisions in the jet of a hadronic blazar with the characteristics of 1ES 1959+650.



\section{Inferring a Neutrino Spectrum From Gamma-Ray Observations}

Assuming the TeV emission observed in a blazar's spectrum is generated in the decay of pions produced in the collisions of protons accelerated in its jets, high energy neutrinos must also be present. Furthermore, the gamma-ray and neutrino fluxes can be related by energy considerations \cite{alvarez}:
\begin{equation}
\int^{E_{\gamma}^{\rm{max}}}_{E_{\gamma}^{\rm{min}}} E_{\gamma} \frac{dN_{\gamma}}{dE_{\gamma}} dE_{\gamma} = K \int^{E_{\nu}^{\rm{max}}}_{E_{\nu}^{\rm{min}}} E_{\nu} \frac{dN_{\nu}}{dE_{\nu}} dE_{\nu},
\label{compare}
\end{equation}
where $K$ is a factor which depends on whether the pions are generated in $pp$ ($K=1$) or $p \gamma$ ($K=4$) collisions.

The spectrum of protons generated via Fermi acceleration in the blazar jets is expected to be of the form $dN_p/dE_p \propto E_p^{-2}$, with both the gamma-ray and neutrino spectra tracing this slope (assuming a slowly varying spectrum for the target photons/protons). Gamma-rays can interact in the source, as well as with the Infra-Red Background (IRB) during propagation, however, steepening their observed spectrum. The threshold for the process, $\gamma + \gamma_{\rm{IRB}} \rightarrow e^+ e^-$, is given by:
\begin{equation}
E_{\gamma} > \frac{2 m_e^2}{E_{\rm{IRB}} (1-\cos \theta) } \sim 10-50 \, \,\rm{TeV}.
\end{equation}
Here, $E_{\rm{IRB}} \sim 0.01$ eV is the energy of a typical IRB photon and $\theta$ is the angle between the two photons. The precise behavior of this interaction on the spectrum of high energy photons depends on the poorly measured distribution of extragalactic infrared radiation, as well as on the distance to the source.

Within the source itself, gamma-ray cascading can also significantly modify the observed spectrum. These processes make it impossible to match the shape of the neutrino spectrum to that of the gamma-rays. The total energy between photons and neutrinos can still be related, however. Even this only yields a {\it lower limit} on the neutrino flux, however, as absorbed photons are not accounted for in the left hand side of Eq.~\ref{compare}.

\subsection{Proton-Proton Collisions}

Following simple relativistic kinematics, for the center-of-mass energy of a proton-proton collision to exceed the energy threshold for pion production, the accelerated proton must have an energy above:
\begin{equation}
E_{p}^{\rm{min}} = \Gamma \, \frac{(2 m_p + m_{\pi})^2 -2 m_p^2 }{2 m_p} \simeq \Gamma \times  1.23 \, \rm{GeV},
\end{equation}
where $\Gamma$ is the Lorentz factor of the jet relative to the observer. The maximum energy to which protons are accelerated depends on the kinematics in the jet. The maximum and minimum gamma-ray and neutrino energies are related to this quantities by:
\begin{equation}
E_{\gamma}^{\rm{max}}=\frac{E_p^{\rm{max}}}{6}, \,\,\, E_{\nu}^{\rm{max}}=\frac{E_p^{\rm{max}}}{12},
\end{equation}
with analogous expressions relating the minimum energies. The factors of 6 and 12 come from, on average, three pions being produced and each charged (neutral) pion decaying into four (two) particles. 

\subsection{Proton-Photon Collisions}

Pions can also be generated in $p \gamma$ collisions via the delta hadron resonance, $p \gamma \rightarrow \Delta \rightarrow \pi N$. For this process to take place, the center-of-mass energy of the interaction must exceed the $\Delta$-mass, 1.232 GeV. This correspond to an energy of:
\begin{equation}
E_{p}^{\rm{min}} = \Gamma^2 \, \frac{m_{\Delta}^2 - m_p^2}{4 E_{\gamma}} \simeq \Gamma^2 \, \bigg(\frac{1 \,\rm{MeV}}{E_{\gamma}}\bigg)  \times 160 \,\rm{GeV}.
\end{equation}
This is clearly a much higher energy cutoff than in the case of proton-proton collisions. To generate TeV photons by this mechanism, radiation clouds surrounding the blazar must contain photons of $\sim$MeV energies \cite{Atoyan:2001ey,Dermer:2001xd,Mucke:2001qg,Protheroe:1996uu}. Such clouds are known to exist around at least some line-emitting blazars. 

In the case of $p \gamma$ collisions, the maximum and minimum neutrino and gamma-ray energies are given by:
\begin{equation}
E_{\gamma}^{\rm{max}}=\frac{E_p^{\rm{max}} <x_{p \rightarrow \pi}> }{2}, \,\,\, E_{\nu}^{\rm{max}}=\frac{E_p^{\rm{max}} <x_{p \rightarrow \pi}>}{4},
\end{equation}
again with analogous relations for minimum energies. $<x_{p \rightarrow \pi}>\simeq 0.2$ is the average fraction of the proton's energy which is transfered to the pion.

\section{Estimates of High Energy Neutrino Fluxes}

Consider a TeV blazar with the observed spectrum:
\begin{equation}
\frac{dN_{\gamma}}{dE_{\gamma}} = A_{\gamma} E_{\gamma}^{- \alpha}.
\label{gammaspec}
\end{equation}
The expectation for the gamma-ray spectrum from Fermi accelerated proton interactions is $\alpha \approx 2$, although the observed spectrum is likely to be steepened due to cascading. We will assume that the neutrino spectrum more closely mimics the Fermi accelerated proton form:
\begin{equation}
\frac{dN_{\nu}}{dE_{\nu}} = A_{\nu} E_{\nu}^{-2}.
\end{equation}
Using Eq.~\ref{compare}, we can relate these coefficients:
\begin{equation}
A_{\nu} \approx \frac{A_{\gamma} E_{\gamma, \rm{min}}^{-\alpha+2}}{(\alpha-2) K \ln{(E_{\nu,\rm{max}}/E_{\nu,\rm{min}})}},
\end{equation}
where we have assumed that $E_{\gamma, \rm{max}} \gg E_{\gamma, \rm{min}}$. The neutrino spectrum from a hadronic TeV blazar with the spectrum of Eq.~\ref{gammaspec} is thus given by:
\begin{equation}
\frac{dN_{\nu}}{dE_{\nu}}   \approx  A_{\nu}  \, E_{\nu}^{-2} \approx \frac{A_{\gamma} E_{\gamma, \rm{min}}^{-\alpha+2}}{(\alpha-2) K \ln{(E_{\nu,\rm{max}}/E_{\nu,\rm{min}})}}  \, E_{\nu}^{-2},
\end{equation}
which for proton-proton collisions is approximately:
\begin{equation}
\frac{dN_{\nu}}{dE_{\nu}} \, E_{\nu}^2 \sim  2 \times 10^{-12}\,\, \rm{TeV} \, \rm{cm}^{-2} \, \rm{s}^{-1} \, \times    \bigg(\frac{A_{\gamma}}{3.2 \times 10^{-11} \, \rm{TeV} \, \rm{cm}^{-2} \, \rm{s}^{-1}}     \bigg) \bigg(\frac{10}{\Gamma}\bigg)^{\alpha-2} \bigg(\frac{2.8-2}{\alpha-2} \bigg),
\label{ppspec}
\end{equation}
and for proton-photon collisions is:
\begin{equation}
\frac{dN_{\nu}}{dE_{\nu}} \, E_{\nu}^2 \sim 4 \times  10^{-13}\,\, \rm{TeV} \, \rm{cm}^{-2} \, \rm{s}^{-1}    \, \times    \bigg(\frac{A_{\gamma}}{3.2 \times 10^{-11} \, \rm{TeV} \, \rm{cm}^{-2} \, \rm{s}^{-1}} \bigg) \bigg(\frac{10}{\Gamma}\bigg)^{2\alpha-4} \bigg(\frac{2.8-2}{\alpha-2} \bigg) \bigg(\frac{E_{\gamma,\rm{target}}}{1\,\rm{MeV}} \bigg)^{\alpha-2}.
\label{pgammaspec}
\end{equation}
In these expressions, $\Gamma$ is the Lorentz factor of the jet. The gamma-ray flux is normalized to that of the Crab nebula: $dN_{\gamma}/dE_{\gamma} \approx 3.2 \times 10^{-11} \, \rm{TeV} \, \rm{cm}^{-2} \, \rm{s}^{-1}$ at 1 TeV.

At the source, two thirds of the neutrinos produced are of muon flavor (the remaining being electron neutrinos). After oscillations which occur during propagation, the neutrino flux reaches Earth in roughly equal fractions of electron, muon and tau flavors.

\section{Event Rate Estimates}

To detect a (muon) neutrino of GeV-TeV energy in a high energy neutrino telescope, an energetic muon must be produced in a charged current interaction near the detector. Once generated, the muon loses energy at a rate:
\begin{equation}
\frac{dE_{\mu}}{dX} \approx -A -B E_{\mu},
\end{equation}
where the values of $A$ and $B$ depend on the medium. For ice, $A \approx 2.0 \times 10^{-3}$ GeV cm$^{-1}$ and $B \approx 4.2 \times 10^{-6}$ cm$^{-1}$. The distance a muon of an energy, $E_{\mu}$, travels before its energy drops below an energy threshold, $E_{\mu}^{\rm{thr}}$, is then given by \cite{dutta}:
\begin{equation}
R_{\mu}(E_{\mu}) \approx \frac{1}{B} \ln \bigg[\frac{A+BE_{\mu}}{A+BE_{\mu}^{\rm{thr}}}\bigg].
\end{equation}
This quantity is called the muon range. The number of neutrino-induced muons observed in a high energy neutrino telescope is given by:
\begin{equation}
N_{\rm{events}} \approx \int \frac{dN_{\nu_{\mu}}}{dE_{\nu_{\mu}}} P_{\nu \rightarrow \mu} A_{\rm{eff}} T dE_{\nu_{\mu}},
\end{equation}
where $A_{\rm{eff}}$ is the effective muon area of the detector, $T$ is the length of time observed (or the duration of a flare) and $P_{\nu \rightarrow \mu}$ is the neutrino to muon transition probability, given by:
\begin{equation}
P_{\nu \rightarrow \mu} \approx \sigma_{\rm{CC},\nu N} R_{\mu} n,
\end{equation}
where $\sigma_{\rm{CC},\nu N}$ is the charged current neutrino-nucleon cross section, $R_{\mu}$ is the muon range and $n$ is the number density of nucleons in the target material.

For a specific experiment, the event rates predicted are determined by sophisticated monte carlos which take into account the detailed geometry and other characteristics of the detector. For the AMANDA detector, the effective neutrino area for a northern hemisphere source is about 1, 100 and 3000 square centimeters for neutrinos of 100 GeV, 1 TeV and 10 TeV energies, respectively. This effective neutrino volume function, convolved with the incoming neutrino spectrum, produces the observed number of events. 

\begin{figure}[t]
\centering\leavevmode
\includegraphics[width=3.3in,angle=90]{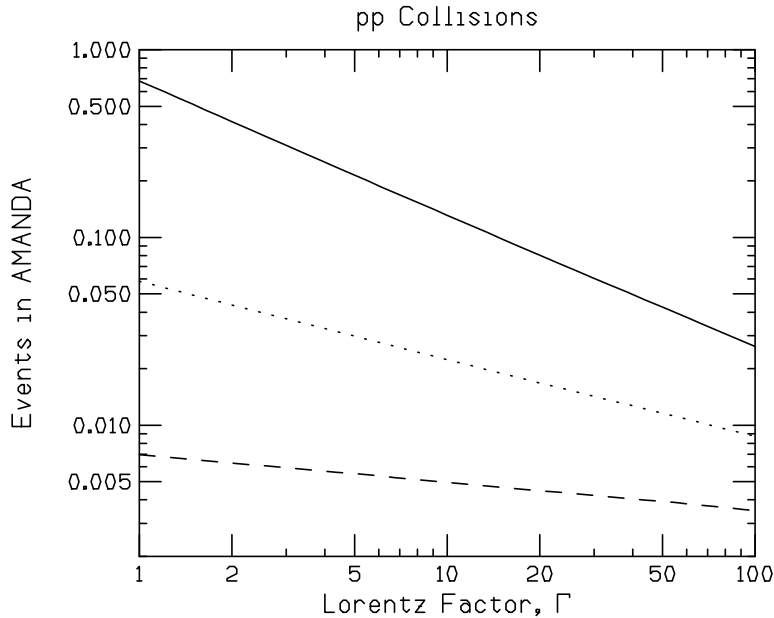}
\caption{The {\it lower limit} on the number of events in AMANDA predicted from a hadronic blazar flare which produces TeV gamma-rays via proton-proton collisions as a function of the Lorentz factor of the jet relative to the observer. The solid, dotted and dashed lines correspond to gamma-ray spectral indexes, $\alpha$, of 2.2, 2.5 and 2.8, respectively. The results are normalized to a gamma-ray flux integrated above 2 TeV equal to that of the Crab nebula, $F_{\gamma}(E_{\gamma} > 2\, \rm{TeV}) \approx 5.6 \times 10^{-12}$ photons cm$^{-2}$ s$^{-1}$, multiplied by a 90 day flare duration. $E_{\gamma}^{\rm{max}}$ and $E_{\nu}^{\rm{max}}$ were set to 20 and 100 TeV, respectively. Note that the rates have not been increased for absorption of photons on the infrared background, nor for possible absorption of the flux in the source.}
\label{pp}
\end{figure}

\begin{figure}[t]
\centering\leavevmode
\includegraphics[width=3.3in,angle=90]{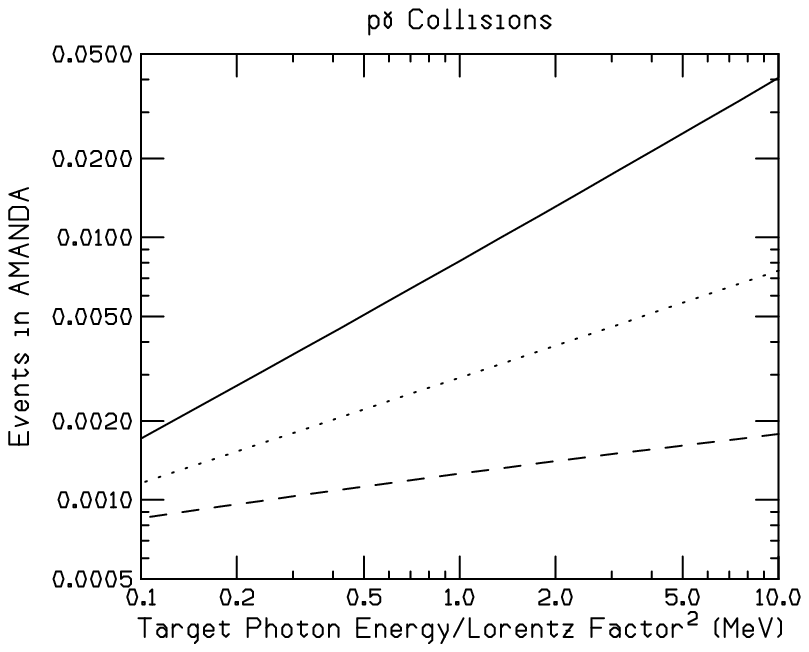}
\caption{The {\it lower limit} on the number of events in AMANDA predicted from a hadronic blazar flare which produces TeV gamma-rays via proton-photon collisions as a function of the energy of the target radiation divided by the Lorentz factor of the jet with respect to the observer squared. The solid, dotted and dashed lines correspond to gamma-ray spectral indexes, $\alpha$, of 2.2, 2.5 and 2.8, respectively. The results are normalized to a gamma-ray flux integrated above 2 TeV equal to that of the Crab nebula, $F_{\gamma}(E_{\gamma} > 2\, \rm{TeV}) \approx 5.6 \times 10^{-12}$ photons cm$^{-2}$ s$^{-1}$, multiplied by a 90 day flare duration. $E_{\gamma}^{\rm{max}}$ and $E_{\nu}^{\rm{max}}$ were set to 20 and 100 TeV, respectively. Note that the rates have not been increased for absorption of photons on the infrared background, nor for possible absorption of the flux in the source.}
\label{pgamma}
\end{figure}

In figures~\ref{pp} and ~\ref{pgamma}, we plot the predicted number of events in AMANDA from hadronic blazar flares for proton-proton and proton-photon collisions, respectively. For each curve plotted, we normalize the rate by setting the gamma-ray flux integrated above 2 TeV to that of the Crab nebula, $F_{\gamma}(E_{\gamma} > 2\, \rm{TeV}) \approx 5.6 \times 10^{-12}$ photons cm$^{-2}$ s$^{-1}$, multiplied by a 90 day exposure (flare duration).

During its flaring activity in 2002, the blazar 1ES 1959+650 was observed by two Atmospheric Cerenkov Telescopes (ACTs), HEGRA and Whipple. The HEGRA collaboration reported a series of flares occurring between May and September of 2002 with integrated fluxes often near or exceeding the flux of the Crab nebula. The brightest of these, which occurred in May, was measured to have an integrated flux above 2 TeV of $1.2 \times 10^{-11}$ photons cm$^{-2}$ s$^{-1}$, about twice the Crab flux. The HEGRA collaboration estimated a spectral index of $2.83 \pm 0.14$ above their threshold of 2 TeV during this activity \cite{hegra}.

The Whipple telescope, operated by the VERITAS collaboration, also reported a series of flares during this period, the brightest of which being almost four times the Crab flux (above 600 GeV). The spectral index reported by Whipple, 2.82 $\pm$ 0.15 $\pm$ 0.3 \cite{daniel}, is in good agreement with the value estimated by HEGRA. 

Inserting these values for the 2002 flaring of 1ES 1959+650 into Eqs.~\ref{ppspec} and \ref{pgammaspec}, we can estimate the neutrino flux and event rates. For proton-proton collisions, using $\alpha=2.83$ and $A_{\gamma}=2 \times 10^{-5}$, we estimate a minimum of 1.8 events observed by AMANDA for a Lorentz factor of $\Gamma =1$ or 0.3 and 0.06 events for $\Gamma =10$ and 100, respectively. The number of events may be considerably higher if a substantial fraction of the energy in gamma-rays is absorbed or cascaded below the energy thresholds of HEGRA and Whipple.

For the case of proton-photon collisions, the neutrino rates are considerably smaller, {\it i.e.} 0.01 for $\Gamma =1$ and MeV target photons. The proton-photon collision scenario appears to be beyond the reach of current detector technology (unless the source is not transparent or absorption on IR background very large), but may be within the sensitivity of next generation experiments, such as IceCube.

\section{Discussion and Conclusions}

The recent observation of neutrinos from the direction of the TeV blazar 1ES 1959+650 coincident with flares occurring in 2002 raises the question of whether such large neutrino fluxes can be generated in these objects. We have found here that such rates are indeed possible for a hadronic blazar which i) accelerates protons which produce pions on nucleonic, rather than photonic, targets, ii) the Lorentz factor of the jet if fairly small ($\Gamma \sim 1$), and iii) the observed spectral index is fairly large. Given the orphan flare observed in 1ES 1959+650, this object is one of the most probable candidates for a gamma-ray source  of hadronic origin. It is also not difficult to imagine a mild boost factor in its jet. Finally, the spectral index of 1ES 1959+650 observed by HEGRA is 2.83, which satisfies the third criterion. 

It is clear that relatively large rates are a consequence of the steeper spectrum for photons relative to neutrinos. One may ask whether the column density in the source required to change a spectral index of -2 into -2.8 is reasonable. Steeper spectra implies higher luminosity given by the left hand side of Eq.\,1. Here again the answer is affirmative. Our highest event rates correspond to luminosities of $10^{47}$-$10^{48}$\,erg/s, not unreasonably high for a flaring blazar. Our calculation also illustrates that where neutrino emission is concerned, not all blazars are alike and that environmental conditions in the jet can be the origin of large variations in the neutrino flux.

\section*{Acknowledgments}

We would like to thank Michael Daniel for informative comments. DH is supported by the Leverhulme trust.

\end{document}